# Supercooled Water: Contradiction to Thermodynamics

I. A. Stepanov

*Institute of Physical Chemistry, Freiburg University*

*Albertstrasse 23a, Freiburg, 79104 Germany*

It has been shown that the dependence of thermal expansion coefficient and of the isobaric heat capacity of supercooled water on the temperature contradicts to an important thermodynamic relation $(\partial C_P/\partial P)_T = -TV((\partial \alpha/\partial T)_P + \alpha^2)$. It has been shown that there is nothing unusual, because the Clapeyron equation also often gives big mistakes despite it is derived from the 1st and 2nd laws of thermodynamics without simplification.

## 1. Introduction

Negative expansivity of supercooled water gets more and more negative as temperature approaches 228 K starting from 273 K. At the same time, the isobaric heat capacity $C_P(T)$ increases [1]. There is a thermodynamic relation using which one can verify the 1st law of thermodynamics [2, 3]:

$$(\partial C_P/\partial P)_T = -TV((\partial \alpha/\partial T)_P + \alpha^2). \qquad (1)$$

Here α is the thermal expansion coefficient. One can show that the dependence $\alpha(T)$ for supercooled water contradicts to Eq. (1).

## 2. Theory

$\alpha^2 \ll |(\partial\alpha/\partial T)_P|$ [1]. In all references where dependence $C_P(P)$ is given, sign$(\partial C_P/\partial P)_T$=sign$(dC_P/dP)$ [4-8]. $C_P=C_P(T, P)$. Let us consider the case that in the 1st approximation $C_P=C_P(P(T))$, in the 2$^{nd}$ approximation $C_P \neq C_P(P(T))$. In the 1st approximation the sign of $dC_P/dP$ is equal the sign of $dC_P/dP$ in the 2$^{nd}$ approximation. But in the 1st approximation,

$$dC_P/dP=(dC_P/dT)dT/dP, \qquad (2)$$

$dT/dP<0$ for $\alpha<0$, for supercooled water $dC_P/dT<0$ and $(\partial\alpha/\partial T)_P>0$ [1]. Therefore, there is a contradiction to Eq. (1). Also, for water with a very big accuracy $C_P(T, P)=C_V(T, P)$ ($C_P \approx 4200$ J/kgK, $C_P-C_V \approx 2.5$ J/kgK [4]) and one can write $dC_P/dP=dC_V/dP=(dC_V/dT)dT/dP$ if $C_V=C_V(P(T))$.

Really, if there is a curve $C_P(T)$ for $P=P_1$ and it is necessary to build a curve $C_P(T)$ for $P=P_1+\Delta P$, it is impossible to build it so that sign$(\partial C_P/\partial P)_T$= -sign$(dC_P/dP)$ in the whole temperature range where $\alpha<0$.

A possible explanation of this phenomenon can be found in [3]. There it has been supposed that for substances with negative thermal expansion, the 1st law of thermodynamics has the following form:

$$\delta Q=dU-PdV. \qquad (3)$$

If to derive Eq. (1) using this formula, one obtains

$$(\partial c_P/\partial P)_T = TV((\partial \alpha/\partial T)_P + \alpha^2). \qquad (4)$$

(It is important to mention that there is a misprint in Eq. (11) in [3]. The correct equation is: for $\alpha<0$, $c_V = c_P k_S/k_T$). In [5] $C_P(P)$ is calculated theoretically, it obeys Eq. (4).

There is a paradox in thermodynamics of water which confirms Eq. (3): $(\partial T/\partial P)_S$ is greater than 0 for $\alpha>0$ and less than 0 for $\alpha<0$ (S is the entropy) [9]:

$$(\partial T/\partial P)_S = (\alpha V/C_P)T \qquad (5)$$

However, $(\partial U/\partial P)_S$ is always greater than 0:

$$(\partial U/\partial P)_S = C_V k_T P V/C_P \qquad (6)$$

The increase in the internal energy dU due to pressure is equal to CdT where C is the adiabatic energy-temperature coefficient $(\partial U/\partial T)_S$, whence temperature decreases when energy increases at $\alpha<0$. It is not true because if one introduces the quantity of heat $\delta Q$ in water at $\alpha<0$, its temperature increases. According to Joule's principle of equivalence of heat and work, it makes no difference whether one increases the internal energy by introducing a quantity of heat or by compression. In my future works it will be shown that $(\partial U/\partial T)_X > 0$ for whatever X. There is a mistake in the traditional thermodynamics that $(\partial U/\partial T)_S < 0$ for $\alpha<0$, this contradiction is intensively discussed in [9].

This contradiction can be explained by Eq. (3). If to derive Eq. (5) using Eq. (3), than for $\alpha<0$,

$$(\partial T/\partial P)_S = -\alpha V/C_P T > 0 \qquad (7)$$

and $(\partial T/\partial P)_S$ is always positive.

In [10] dependences $C_P(P)$ and $\alpha(T)$ are given for $\alpha<0$. One sees clearly that they obey Eq. 4, not Eq. 1. (Take into account that $\alpha(T=0\ °C) > \alpha(T=1\ °C)$).

One has to note that there is a number of paradoxes in thermodynamics of substances with negative expansion [11-17]. For example, the Clapeyron equation

$$dT/dP = \Delta V/\Delta S \qquad (3)$$

gives huge mistakes in describing substances with negative $\alpha$ despite it is derived from the 1st and 2nd laws of thermodynamics without simplification. So, one does not have to surprise that Eq. 1 for some substances is wrong.

In [12, 13, 15-17] polymorphic transitions with increasing coordination are described whence $\Delta S<0$ and $\Delta V<0$, but $dT/dP<0$.